# Personal Information Databases

Sabah S. Al-Fedaghi
Computer Engineering Department
Kuwait University
Kuwait
sabah@eng.kuniv.edu.kw

Bernhard Thalheim
Computer Science Institute
Kiel University
Kiel, Germany
thalheim@is.informatik.uni-kiel.de

*Abstract*—One of the most important aspects of security organization is to establish a framework to identify security-significant points where policies and procedures are declared. The (information) security infrastructure comprises entities, processes, and technology. All are participants in handling information, which is the item that needs to be protected. Privacy and security information technology is a critical and unmet need in the management of personal information. This paper proposes concepts and technologies for management of personal information. Two different types of information can be distinguished: personal information and non-personal information. Personal information can be either personal-identifiable information (PII), or non-identifiable information (NII). Security, policy, and technical requirements can be based on this distinction. At the conceptual level, PII is defined and formalized by propositions over infons (discrete pieces of information) that specify transformations in PII and NII. PII is categorized into simple infons that reflect the proprietor's aspects, relationships with objects, and relationships with other proprietors. The proprietor is the identified person about whom the information is communicated. The paper proposes a database organization that focuses on the PII spheres of proprietors. At the design level, the paper describes databases of personal identifiable information built exclusively for this type of information, with their own conceptual scheme, system management, and physical structure.

*Keywords-component; database; personal information; personal identifiable information*

## I. INTRODUCTION

Rapid advances in information technology and the emergence of privacy-invasive technologies have made information privacy a critical issue. According to Bennett and Raab [11], technically, the concept of information privacy is treated as information security. "Information privacy is the interest an individual has in controlling, or at least significantly influencing, the handling of data *about* themselves" [10]; however, the information privacy domain goes beyond security concerns.

Information security aims to ensure the security of all information regardless whether privacy related or non-privacy related. Here we use the term *information* in its ordinary sense of "facts" stored in a database. This paper explores the privacy-related differences between types of information to argue that security, policy, and technical requirements set personal identifiable information apart from other types of information, leading to the need for a PII database with its own conceptual scheme, system management, and physical structure.

Different types of information of interest in this paper are shown in Fig. 1. We will use the term *infon* to refer to "a piece of information" [9]. The *parameters* of an infon are objects, and so-called *anchors* assign these objects such as agents to parameters. Infons can have sub-infons that are also infons.

Let INF = the set of infons in the system. Four types of infons are identified:

1. So-called "private or personal" information is a subset of INF. "Private or personal" information is partitioned into two types of information: PII and PNI.
2. PII is the set of pieces of personal identifiable information. We use the term *pinfon* to refer to this special type of infon. The relationship between PII and the notion of identifiably will be discussed later.
3. PNI is the set of pieces of non-identifiable information.

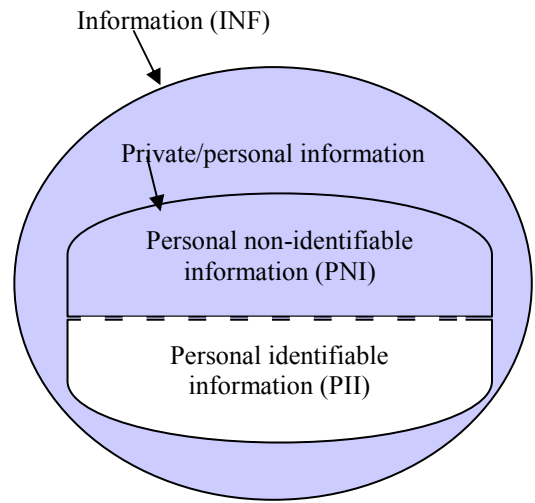

Figure 1.  Types of information.

4. NII = (INF − PII). We use the term *ninfon* to refer to this special type of infon. NII is the set of pieces of

non-identifiable information and includes all pieces of information except personal identifiable information (shaded area in Fig. 1). PNI in Fig. 1 is a subset of NII. It is the set of non-identifiable information; however, it is called "personal" or "private" because its owner (a natural person) has interest in keeping it private. In contrast, PII embeds a unique identity of a natural person

From the security point of view, PII is more *sensitive* than an "equal amount" (to be discussed later) of NII. With regard to policy, PII has more policy-oriented significance (e.g., the 1996 EU directive) than NII. With regard to technology, there are unique PII-related technologies (e.g., P3P) and techniques (e.g., k-anonymization) that revolve around PII. Additionally, PII possesses an objective definition that allows separating it from other types of information, which facilitates organizing it in a manner not available to NII information.

The distinction of infons into PII, NII, and PNI requires a supporting technology. We thus need a framework that allows us to handle, implement, and manage PII, NII, and PNI. Management of PII, NII, and PNI ought, ideally, to be optimal in the sense that derivable infons are not stored. This paper introduces a formalism to specify privacy-related infons based on a theoretical foundation. Current privacy research lacks such formalism. The new formalism can benefit two areas. First, a precise definition of the informational privacy notion is introduced. It can also be used as a base to develop a formal and informal specification language. Informal specification language can be used as a vehicle to specify various privacy constraints and rules. Further work can develop a full formal language to be used in privacy enhancing systems.

In this paper, we concentrate on the *conceptual* organization of PII databases based on a theory of infons. To achieve such a task, we need to identify which subset of infons will be considered personal identifiable information. Since the combination of personal identifiable information is also personally identifiable, we must find a way to minimize the information to be stored. We introduce an algebra that supports such minimization. Infons may belong to different users in a system. We distinguish between *proprietors* (persons to whom PII refers through embedded identities) and owners (entities that possess PII of others such as agencies or other non-proprietor persons).

## II. Related Works

Current database management systems (DBMS) do not distinguish between PII and NII. An enterprise typically has one or several databases. Some data is "private," other data is public, and it is typical that these data are combined in queries. "Private" typically means exclusive ownership of and rights (e.g., access) to the involved data, but there is a difference between "private" data and personal identifiable data. "Private" data may include NII exclusively controlled by its owner; in contrast, PII databases contain only personal identifiable information and related data, as will be described later. For example, in the Oracle database, the Virtual Private Database (VPD) is the aggregation of fine-grained access control in a secure application context. It provides a mechanism for building applications that enforce the security policies customers want enforced, but only where such control is necessary. By dynamically appending a predicate to SQL statements, VPD limits access to data at the table's row level and ties the security policy to the table (or view) itself. "Private" in VPD means data owned and controlled by its owner. Such a mechanism supports the "privacy" of any owned data, not necessarily personal identifiable information. In contrast, we propose to develop a general PII information database management system where PII and NII are explicitly separated in planning, design, and implementation.

Separating "private" data from "public" data has already been adopted in privacy preserving systems; however, these systems do not distinguish explicitly personal identifiable information. The Platform for Privacy Preferences (P3P) is one such system that provides a means for privacy policy specification and exchange but "does not provide any mechanism to ensure that these promises are consistent with the internal data processing" [7]. It is our judgment that "internal data processing" requires recognizing explicitly that "private data" is of two types: personal identifiable information and personal *non-identifiable* information, and this difficulty is caused by the heterogeneity of data. Hippocratic databases have been introduced as systems that integrate privacy protection into relational database systems [1][4]. A Hippocratic database includes privacy policies and authorizations associated with each attribute and each user for usage purpose(s). Access is granted if the access purpose (stated by the user) is entailed by the allowed purposes and not entailed by the prohibited purposes [7]. Users' role hierarchies, similar to ones used in security policies (e.g., RBAC), are used to simplify management of the mapping between users and purposes. A request to access data is accompanied by access purpose, and accessing permission is determined after comparing such purpose with the intended purposes of that data in privacy policies. Each user has authorization for a set of access purposes. Nevertheless, in principle, a Hippocratic database is a general DBMS with a purpose mechanism. Purposes can be declared for any data item that is not necessarily personal identifiable information.

## III. Infons

This section reviews the theory of infons. The theory of infons provides a rich algebra of construction operations that can be applied to PII. Infons in an application domain such as personal identifiable information are typically interrelated; they partially depend on each other, partially exclude each other, and may be (hierarchically) ordered. Thus we need a theory that allows constructing a "lattice" of infons (and PII infons) that includes basic and complex infons while taking into consideration their structures and relationships. In such a theory, we identify basic infons that cannot be decomposed into more basic infons. This construction mechanism of infons from infons should be supported by an algebra of construction operations. We generally may assume that each infon consists of a number of components. The construction is applied in performing combination, replacement, or removal of some of these components; some may be essential (not removable) or auxiliary (optional).

An *infon* is a discrete item of information and may be parametric and anchored. The *parameters* represent objects or properties of objects. *Anchors* assign objects to parameters. Parameter-value pairs are used to represent a property of an infon. The property may be valid, invalid, undetermined, etc. The validity of properties is typically important information. Infons are thus representable by a tuple structure

<<ID, {(param, value, validity)} >>

or by an anchored tuple structures

<<ID, {((param, value, validity), anchor(object))} >>.

We may order properties and anchors. A linear order allows representing an infon as a simple predicate value. Following Devin's formalism [9], an infon has the form <<R, a1, ... , an, 1>> and <<R, a1, ... , an, 0>>. R is an n-place relation and a1, . . . , an are objects appropriate for R. 0 and 1 indicate these may be thought of as objects do, do not, respectively, and they stand in relation R. For simplicity sake, we may write an infon <<R, a1, ... , an, 1/0>> as <<a1, ... , an>> when R is known or immaterial.

We may use multisets instead of sets for infons or a more complex structure. We choose the set notation because of its representability within the XML technology. Sets allow us to introduce a simple algebra and a simple set of predicates.

"PII infons" are distinguished by the mandatory presence of at least one proprietor, an object of type uniquely identifiable person.

The *world of infons* currently of interest can be specified as the triple: (*A*; *O*; *P*) as follows.

- Atomic infons *A*
- Algebraic operations *O* for computing complex infons such as combination ⊕ of infons, abstraction ⊗ of infons by projections, quotient ÷ of infons, ρ renaming of infons, union ∪ of infons, intersection ∩ of infons, full negation ¬ of infons, and minimal negation ¬ of infons within a given context.
- Predicates *P* stating associations among infons such as the sub-infon relation, a statement whether infons can be potentially associated with each other, a statement whether infons cannot be potentially associated with each other, a statement whether infons are potentially compatible with each other, and a statement whether infons are incompatible with each other.

The combination of two infons results in an infon with all components of the two infons. The abstraction is used for a reduction of components of an infon. The quotient allows concentrating on those components that do not appear in the second infon. The union takes all components of two infons and does not combine common components into one component. The full negation allows generating all those components that do not appear in the infon. The minimal negation restricts this negation to some given context.

We require that the sub-infon relation is not transitively reflexive. The compatibility and incompatibility predicates are not contradictory. The potential association and its negation must not conflict. The predicates should not span all possible associations among the infons but only those that are meaningful in a given application area. We may assume that two infons are either potentially associated or cannot be associated with each other. The same restriction can be made for compatibility.

This infon world is very general and allows deriving more advanced operations and predicates. If we assume the completeness of compatibility and association predicates, we may use expressions defined by the operations and derived predicates. The extraction of application-relevant infons from infons is supported by five operations:

1. Infon *projection* narrows the infon to those parts (objects or concepts, axioms or invariants relating entities, functions, events, and behaviors) of concern for the application-relevant infons. For example, a projection operation may produce the set of proprietors from a given infon, e.g., {Mary, John} from *John loves Mary*.
2. Infon *instantiation* lifts the general infons to those of interest within the solution and instantiates variables by values that are fixed for the given system. For example a PII infon may be instantiated from its anonymized version, e.g., *John is sick* from *Someone is sick*.
3. Infon *determination* is used to select those traces or solutions to the problem under inspection that are the most suitable or the best fit for the system envisioned. The determination typically results in a small number of scenarios for the infons to be supported, for example, infon determination to decide whether an infon belongs to a certain *piiSphere* (PII of a certain proprietor – to be discussed later).
4. Infon *extension* is used to add those facets not given by the infon but by the environment or the platforms that might be chosen or that might be used for simplification or support of the infon (e.g., additional data, auxiliary functionality), for example, infon *extension* to related non-identifiable information (to be discussed later).
5. Infons are often associated, adjacent, interacting, or fit with each other. *Infon join* is used to combine infons into more complex and combined infons that describe a complex solution, for example, joining *atomic* PIIs to form *compound* PII (these types of PII will be defined later) and a collection of related PII information.

The application of these operations allows extraction of which sub-infons, which functionality, which events, and which behavior (e.g., the action/verb in PII) are *shared* among information spheres (e.g., of proprietors). These shared "facilities" encompass all information spheres of relevant infons. They also hint at possible architectures of information and database systems and at separation into candidate components. For instance, *entity sharing* (say, non-person entity) describes which information flow and development can be observed in the information spheres.

We will not be strictly formal in applying infon theory to PII. Such a venture needs far more space. Additionally, we squeeze the approach in the area of database design in order to illustrate a sample application. The theory of PII infons can be applied in several areas, including the technical and legal aspects of information privacy and security.

## IV. Personal Identifiable Information

It is typically claimed that what makes data "private" or "personal" is either specific legislation, e.g., a company must not disclose information about its employees, or individual agreements, e.g., a customer has agreed to an electronic retailer's privacy policy. However, this line of thought blurs the difference between personal identifiable information and other "private" or "personal" information. Personal identifiable information has an "objective" definition in the sense that it is independent of such authorities as legislation or agreement.

PII infons involve a special relationship called *proprietorship* with their proprietors, but not with persons who are their non-proprietors, and non-persons such as institutions, agencies, or companies. For example, a person may possess PII of another person, or a company may have the PII of someone in its database; however, *proprietorship* of PII is reserved only to its proprietor regardless of who possesses it.

To base personal identifiable information on firmer ground, we turn to stating some principles related to such information. For us, personal identifiable information (*pinfon*) is any information that has referent(s) to uniquely identifiable persons [2]. In logic, this type of *reference* is the relation of a word (logical name) to a thing.

A pinfon is an infon such that at least one of the "objects" is a singly identifiable person. Any singly identifiable person in the pinfon is called *proprietor* of that pinfon. The proprietor is the person about whom the pinfon communicates information. If there is exactly one object of this type, the pinfon is an *atomic* pinfon; if there is more than one singly identifiable person, it is a *compound* pinfon. An atomic pinfon is a discrete piece of information about a singly identifiable person. A *compound* pinfon is a discrete piece of information about several singly identifiable persons. If the infon does not include a singly identifiable person, it is called a *ninfon*.

We now introduce a series of axioms that establish the foundation of the theory of personal identifiable information. These axioms can be considered negotiable assumptions. The symbol "→" denotes implication. INF is the set of infons described in Fig. 1.

### 1. Inclusivity of INF

$\sigma \in INF \leftrightarrow \sigma \in PII \lor \sigma \in NII$

That is, infons are the union of pinfons and ninfons. PII is the set of pinfons (pieces of personal identifiable information), and NII is the set of ninfons (pieces of non-identifiable information).

### 2. Exclusivity of PII and NII

$\sigma \in INF \land \sigma \notin PII \rightarrow \sigma \in N$
$\sigma \in INF \land \sigma \notin N \rightarrow \sigma \in PII$

That is, every infon is exclusively either pinfon or ninfon.

### 3. Identifiability

Let ID denote the set of (basic) pinfons of type
$<< is, Þ, 1>>$ and let þ be a parameter for a singly identifiable person.
Then $<< is, Þ, 1>> \rightarrow << is, Þ, 1>> \in INF$

### 4. Inclusivity of PII

Let $n_\sigma$ denote the number of uniquely identified persons in the infon $\sigma$, then $\sigma \in INF \land n\sigma > 0 \leftrightarrow \sigma \in PII$

### 5. Proprietary

For $\sigma \in PII$, let $PROP(\sigma)$ be the set of proprietors of $\sigma$. Let PERSONS denote the set of (natural) persons. Then,
$\sigma \in PII \rightarrow PROP(\sigma) \in PERSONS$
That is, pinfons are pieces of information about persons.

### 6. Inclusivity of NII

$\sigma \in INF \land (n_\sigma = 0) \leftrightarrow \sigma \in NII$

That is, non-identifiable information (ninfon) does not embed any unique identifiers of persons.

### 7. Combination of non-identifiability with identity

Let ID denote the set of (basic) pinfons of type:

$<< is, Þ, 1>>$, then,
$\sigma_1 \in PII \leftrightarrow <<\sigma_2 \in NII \oplus \sigma_3 \in ID) >>$
assuming $\sigma_1 \notin ID$. "$\oplus$" here denotes the "merging" of two sub-infons.

### 8. Closability of PII

$\sigma1 \in PII \oplus \sigma2 \in PII \rightarrow (\sigma1 \oplus \sigma2) \in PII$

### 9. Combination with non-identifiability

$\sigma_1 \in NII \oplus \sigma_2 \in PII \rightarrow (\sigma_1 \oplus \sigma_2) \in PII$

That is, non-identifying information plus personal identifiable information is personal identifiable information.

### 10. Reducibility to non-identifiability

$\sigma_1 \in PII \div (\sigma_2 \in ID) \leftrightarrow \sigma_3 \in NII$
where $\sigma2$ is a sub-infon of $\sigma1$. "$\div$" denotes removing $\sigma2$.

### 11. Atomicity

Let APII = the set of *atomic* personal identifiable information. Then, $\sigma \in PII \land (n\sigma = 1) \leftrightarrow \sigma \in APII$

### 12. Non-atomicity

Let CPII = the set of *compound* personal identifiable information. Then, $\sigma \in PII \land (n\sigma > 1) \leftrightarrow \sigma \in CPII$

### 13. Reducibility to atomicity

$\sigma \in CPII \leftrightarrow <<\sigma_1, \sigma_2, …, \sigma_m >>$, $\sigma i \in APII$,

$m = n_\sigma$, and $1 \leq i \leq m$, and $\{PROP(\sigma_1), PROP(\sigma_2), …, PROP(\sigma_m)\} = PROP(\sigma)$.

These axioms support the separation of infons into PII, NII, and PNI and their transformation. Let us now discuss the impact of some of these axioms. We concentrate the discussion on the more difficult axioms.

### Identifiability

Let þ be a parameter for a singly identifiable person, i.e., a specific person, defined as

Þ = IND₁|<< singly identifiable, IND₁, 1>>

where IND indicates the basic type: an individual [9].

That is, Þ is a (restricted) parameter with an anchor for an object of type singly identifiable individual. The individual IND₁ is of type person defined as

<< person, IND₁, 1>>

Put simply, þ is a *reference* to a singly identifiable person. We now elaborate on the meaning of "identifiable."

Consider the set of unique *identifiers* of persons. Ontologically, the Aristotelian entity/object is a single, specific existence (a particularity) in the world. For us, the identity of an entity is its *natural descriptors* (e.g., tall, black eyes, male, blood type A, etc.). These descriptors *exist in* the entity/object. Tallness, whiteness, location, etc. exist as aspects of the existence of the entity. We recognize the human entity from its natural descriptors. Some descriptors form *identifiers*. A *natural identifier* is a set of natural descriptors that facilitates recognizing a person *uniquely*. Examples of identifiers include fingerprints, faces, and DNA. No two persons have identical natural identifiers. An *artificial descriptor* is a descriptor that is mapped to a natural identifier. Attaching the number 123456 to a particular person is an example of an artificial descriptor in the sense that it is not recognizable in the (natural) person. *An artificial identifier* is a set of descriptors mapped to a natural identifier of a person. Date of birth (an artificial descriptor), gender (a natural descriptor), and a 5-digit ZIP (an artificial descriptor) are three descriptors that form an artificial identifier for 87% of the US population [12]. By implication, no two persons have identical artificial identifiers. If two persons somehow have the same Social Security number, then this Social Security number is not an artificial identifier because it is not mapped uniquely to a natural identifier.

We define identifiers of proprietors as infons. Such definition is reasonable since the mere act of identifying a proprietor is a *reference* to a unique entity in the information sphere. Hence,

<< is, Þ, 1>> → << is, Þ, 1>> ∈ INF

That is, every unique identifier of a person is an infon. These infons cannot be decomposed into more basic infons.

**Inclusivity of PII**

Next we position identifiers as the basic infons in the sphere of PII. The symbol n_σ denotes the number of uniquely identified persons in infon σ. Then we can define PII and NII accordingly:

σ ∈ INF ∧ n_σ > 0 ↔ σ ∈ PII

That is, an infon that includes unique identifiers of (natural) persons is personal identifiable information. From (3) and (4), <u>any unique personal identifier or piece of information that embeds identifiers is personal identifiable information</u>. Thus, identifiers are the basic PII infons (pinfons) that cannot be decomposed into more basic infons. Furthermore, every complex *pinfon* includes in its structure at least one basic infon, i.e., identifier. The structure of a *complex* pinfon is constructed from several components:

- Basic pinfons and ninfons, i.e., the pinfon *John S. Smith* and the ninfon *Someone is sick* form the atomic PII (i.e., PII with one proprietor) *John S. Smith is sick*. This pinfon is produced by an *instantiation* operation that lifts the general infons to pinfons and instantiates the variable (*Someone*) by a value (*John S. Smith*).
- Complex pinfons form more complex infons, e.g., *John S. Smith and Mary F. Fox are sick*

We notice that the operation of *projection* is not PII-closed since we can define projecting of ninfon from pinfon (removing all identifiers). This operation is typically called anonymization.

Every pinfon refers to its *proprietor(s)* in the sense that it "leads" to him/her/them as distinguishable entities in the world. This reference is based on his/her/their unique identifier(s). As stated previously, the relationship between persons and their own pinfon is called *proprietorship* [1]. A pinfon is proprietary PII of its *proprietor*(s).

Defining pinfon as "information identifiable to the individual" does not mean that the information is "especially sensitive, private, or embarrassing. Rather, it describes a relationship between the information and a person, namely that the information—whether sensitive or trivial—is somehow identifiable to an individual" [10]. However, personal identifiable information (pinfon) is more "valuable" than personal non-identifiable information (ninfon) because it has an intrinsic value as "a human matter," just as privacy is a human trait. Does this mean that scientific information about how to make a nuclear bomb has less intrinsic moral value than the pinfon *John is left handed*? No, it means *John is left handed* has a higher moral value than the ninfon *There exists someone who is left handed*. It is important to compare equal amounts of information when we decide the status of each type of information [5].

To exclude such notions as confidentiality being applicable to the informational privacy of non-natural persons (e.g., companies), the next axiom formalizes that pinfon is applied only to (natural) persons.

For σ ∈ PII, we define PROP(σ) to be the set of proprietors of σ. Notice that |PROP(σ ∈ PII)| = n_σ. Multiple occurrences of identifiers of the same proprietor are counted as a single reference to the proprietor. In our ontology, we categorize things (in the world) as *objects* (denoted by the set OBJECTS) and *non-objects*. Objects are divided into (natural) persons (denoted by the set PERSONS) and non-persons. A fundamental proposition in our system is that proprietors are (natural) persons.

**Combination of non-identifiability with identity**

Next we can specify several transformation rules that convert from one type of information to another. These (privacy) rules are important for deciding what type of information applies to what operations (e.g., information disclosure rules).

Let ID denote the set of (basic) pinfons of type << is, Þ, 1>>. That is, ID is the set of identifiers of persons (in the world). We now define construction of complex infons from basic pinfons and non-identifying information. The definition also applies to *projecting* pinfons from more complex pinfons by removing all or some non-identifying information.

$$\sigma_1 \in PII \leftrightarrow <<\sigma_2 \in NII \oplus \sigma_3 \in ID) >>$$

assuming $\sigma_1 \notin ID$.

That is, non-identifiable information plus a unique personal identifier is personal identifiable information and vice versa (i.e., minus). Thus the set of pinfons is closed under operations that remove or add non-identifying information. We assume the empty information $\varnothing$ is in NII. "$\oplus$" here denotes "merging" two sub-infons. We also assume that only a single $\sigma_3 \in ID$ is added to $\sigma_2 \in NII$; however, the axiom can be generalized to apply to multiple identifiers. An example of axiom 7 is

$\sigma_1 = <<$ *John loves apples*$>> \leftrightarrow <<\sigma_2 =$ *Someone loves apples* $\oplus \sigma_3 =$ *John*$>>$

Or, in a simpler description: $\sigma_1 =$ *John loves apples* $\leftrightarrow \{\sigma_2 =$ *Someone loves apples* $\oplus \sigma_3 =$ *John*$\}$

The axiom can also be applied to the union $\cup$ of pinfons.

**Closability of PII**

PII is a closed set under different operations (e.g., merge, concatenate, submerge, etc.) that construct complex pinfons from more basic pinfons. Hence,

$$\sigma_1 \in PII \oplus \sigma_2 \in PII \rightarrow (\sigma_1 \oplus \sigma_2) \in PII$$

That is, merging personal identifiable information with personal identifiable information produces personal identifiable information. In addition, PII is a closed set under different operations (e.g., merge, concatenate, submerge, etc.) that construct complex pinfons by mixing pinfons with non-identifying information.

**Reducibility to non-identifiability**

Identifiers are the basic pinfons. Removing all identifiers from a pinfon converts it to non-identifying information. Adding identifiers to any piece of non-identifying information converts it to a pinfon,

$$\sigma_1 \in PII \div \sigma_2 \in ID \leftrightarrow \sigma_3 \in NII$$

where $\sigma_2$ is a sub-infon of $\sigma_1$.

Axiom 10 states that personal identifiable information minus a unique personal identifier is non-identifying information and vice versa. "$\div$" here denotes removing $\sigma_2$. We assume that a single $\sigma_2 \in ID$ is embedded in $\sigma_1$; however, the opposition can be generalized to apply to multiple identifiers such that removing all identifiers produces $\sigma 3 \in NII$.

**Atomicity**

Furthermore, we define atomic and non-atomic (compound) types of pinfons. Let
APII = a set of *atomic* personal identifiable information.
Each piece of atomic personal identifiable information is a special type of pinfon called *apinfon*.
As we will see later, *cpinfons* can be reduced to *apinfons*, thus simplifying the analysis of PII. Formally, the set APII is defined as follows. $\sigma \in PII \wedge n_\sigma = 1 \leftrightarrow \sigma \in APII$

That is, an *apinfon* is a pinfon with a single human referent. Notice that $\sigma$ may embed several identifiers of the same person, yet the referent is still one. Notice that *apinfons* can be basic (a single identifier) or complex (a single identifier plus non-identifiable information).

**Non-atomicity**

Let CPII = a set of *compound* personal identifiable information. Each piece of compound personal identifiable information is a special type of pinfon called *cpinfon*. Formally, the set CPII is defined as follows.

$$\sigma \in PII \wedge n_\sigma > 1 \leftrightarrow \sigma \in CPII$$

That is, a cpinfon is a pinfon with more than one human referent. Notice that *cpinfons* are always complex since they must have at least two apinfons (two identifiers).

The apinfon (*atomic personal identifiable information*) is the "unit" of personal identifiable information. It includes one identifier and non-identifiable information. We assume that at least some of the non-identifiable information is *about* the proprietor. In theory this is not necessary. Suppose that an identifier is amended to a random piece of non-identifiable information (noise). In the PII theory the result is (complex) atomic PII. In general, mixing noise with information preserves information.

**Reducibility to atomicity**

Any cpinfon is privacy-reducible to a set of apinfons (atomic personal identifiable information). For example, *John and Mary are in love* can be privacy-reducible to the apinfons *John and someone are in love* and *Someone and Mary are in love*. Notice that our PII theory is a syntax (structural) based theory. It is obvious that the privacy-reducibility of compound personal identifiable information causes a loss of "semantic equivalence," since the identities of the referents in the original information are separated. Semantic equivalency here means preserving the totality of information, the pieces of atomic information, and their link.

Privacy reducibility is expressed by the following axiom:
$\sigma \in CPII \leftrightarrow <<\sigma_1, \sigma_2, …, \sigma_m >>$, $\sigma i \in APII$,

$m = n_\sigma$, $(1 \leq i \leq m)$, and $\{PROP(\sigma_1), PROP(\sigma_2), …, PROP(\sigma_m)\} = PROP(\sigma)$.

The reduction process produces m atomic personal identifiable information with m different proprietors. Notice that the *set* of resultant apinfons produces a compound pinfon. This preserves the totality of the original cpinfon through linking its apinfons together as members of the same set.

V. CATEGORIZATION OF ATOMIC PERSONAL IDENTIFIABLE INFORMATION

In this section, we identify categories of apinfons. Atomic personal identifiable information provides a foundation for structuring pinfons since compound personal identifiable information can be reduced to a set of apinfons. We concentrate on reducing all given personal identifiable information to sets of apinfons. Justification for this will be discussed later.

*A. Eliminating ninfons embedded in an apinfon*

Organizing a database of personal identifiable information requires filtering and simplifying apinfons to more basic apinfons in order to make the structuring of pinfons easier. Axiom (9) tells us that pinfons may carry non-identifiable information, ninfons. This non-identifiable information may be random noise or information not directly about the proprietor. Removing random noise is certainly an advantage in designing a database. Identifying information that is not about the proprietor clarifies the boundary between PII and NII.

A first concern when analyzing an apinfon is projecting (isolating, factoring) information *about* any other entities besides the proprietor. Consider the apinfon *John's car is fast*. This is information about John and about a car of his. This apinfon can be projected as:

$$\otimes (\textit{John's car is fast}) \Rightarrow \{\textit{The car is fast}, \textit{John has a car}\},$$
where $\Rightarrow$ is a production operator.

*John's car is fast* information embeds the "pure" apinfon *John has a car* and the ninfon *The car is fast*. *John has a car* is information about a relationship that John has with another object in the world. This last information is an example of what we call *self information*. Self information (sapinfon = self atomic pinfon) is information about a proprietor, his/her aspects (e.g., tall, short), or his/her relationship with non-human objects in the world; it is thus useful to further reduce apinfons (atomic) to sapinfon (self).

Sapinfon is related to the concept of "what the piece of apinfon is about." In the theory of *aboutness*, this question is answered by studying the text structure and assumptions of the source about the receiver (e.g., reader). We formalize *aboutness* in terms of the procedure $ABOUT(\sigma)$, which produces the set of entities/objects that $\sigma$ is "talking" about. In our case, we aim to reduce any self infon $\sigma$ to $\sigma'$ such that $ABOUT(\sigma)$ is $PROP(\sigma')$.

Self atomic information represents information about the following:

- *Aspects* of proprietor (identification, character, acts, etc.)

- His or her association with non-person "things" (e.g., house, dog, organization, etc.)

- His or her relationships with other persons (e.g., *Smith saw a blond woman*).

With regard to non-objects, of special importance for privacy analysis are *aspects* of persons that are expressed by sapinfon. Aspects of a person include his/her (physical) parts, character, acts, condition, name, health, race, handwriting, blood type, manner, and intelligence. The existence of these aspects depends on the person, in contrast to (physical or social) objects associated with him/her such as his/her house, dog, spouse, job, professional associations, etc.

Let SAPII denote the set of sapinfons (self personal identifiable information).

**14. Aboutness proposition**

$$\sigma \in \text{SAPII} \leftrightarrow \text{ABOUT}(\sigma) = \text{PROP}(\sigma)$$

That is, atomic personal identifiable information $\sigma$ is said to be *self* personal identifiable information (sapinfon) if its subject is its proprietor. The term "subject" here means what the entity is *about* when the information is communicated. The mechanism (e.g., manually) that converts APII to SAPII has yet to be investigated.

*A. Sapinfons involving aspects of proprietor or relationship with non-person*

We further simplify sapinfons. Let $OPJ(\sigma \in \text{SAPII})$ be the set of objects in $\sigma$. SAPII is of two types depending on the number of objects embedded in it: singleton, *ssapinfon* and multitude, *msapinfon*. The set *ssapinfons*, SSAPII, is defined as:

**15. Singleton proposition**

$$\sigma \in \text{SSAPII} \rightarrow \sigma \in \text{SAPII} \wedge (\text{PROP}(\sigma) = \text{OPJ}(\sigma))$$

That is, the proprietor of $\sigma$ is its only object.

The set *msapinfons*, MSAPII, is defined as follows.

**16. Multitude proposition**

$$\sigma \in \text{MSAPII} \rightarrow \sigma \in \text{SAPII} \wedge (|\text{OPJ}(\sigma)| > 1)$$

That is, $\sigma$ embeds other objects beside its proprietor.
We also assume logical simplification that eliminates conjunctions and disjunctions of SAPII [5].

Now we can declare that the sphere of personal identifiable information (piiSphere) for a given proprietor is the database that contains:

1. All *ssapinfons* and *msapinfons* of the proprietor, including their arrangement in super-infons (e.g., to preserve compound personal identifiable information).

2. Related non-identifiable information to the piiSphere of the proprietor, as discussed next.

*A. What is related non-identifiable information?*

Consider the msapinfons *Alice visited clinic Y*. It is msapinfons because it represents a relationship (not an aspect of) the proprietor Alice had with an object, the clinic. Information *about* the clinic may or may not be privacy related information. For example, year of opening, number of beds, and other information about the clinic are not privacy related information; thus, such information ought not be included in Alice's piiSphere. However, when the information is that the clinic is an abortion clinic, then Alice's piiSphere ought to include this non-identifiable information about the clinic.

As another example in terms of database tables, consider the following three tables representing the database of a company:

Customer (Civil ID, Name, Address, Product ID)

Product (ID, Price, Factory)
Factory (Product ID, Product location, Inventory)

Customer's piiSphere includes:

- Ssapinfons (aspects of customer): Civil ID, Name
- Msapinfons (relationships with non-person objects): Address, Product ID

However, information about Factory is not information related to the customer's piiSphere.
Now suppose that we have the following database:

Person (Name, Address, Place of work)
Place of work (Name, Owner)

If it is known that the owner of the place of work is the Mafia, then the information related to the person's piiSphere extends beyond the name of place of work.

The decision about the boundary between a certain piiSphere and its related non-identifiable information is difficult to formalize. Fig. 2 shows a conceptualization of piiSpheres of two proprietors that have compound PII. Dark circles A–G represent possible non-person objects. For example, object A participates in an msapinfon (e.g., Factory, Address, and Place of work in previous examples). Object A has its own aspects (white circle around A) and relationships (e.g., with F) where some information may be privacy-significant to the piiSphere of proprietor.

Even the relationship between the two proprietors may have its own sphere of information (white circle around E). E signifies a connection among a set of apinfons (atomic PII) since we assume that all compound PII have been reduced to atomic PII. For example, the infon {*Alice is the mother of a child in the orphanage*, *John is the child of a woman who gave him up*} is a cpinfon with two apinfons. If Alice and John are the proprietors, then E in the figure preserves the connection between these two apinfons in the two piiSpheres of Alice and John.

VI. JUSTIFICATIONS FOR PII DATABASES

We concentrate on what we call PII database, PIIDB, that contains personal identifiable information and information related to it.

*A. Security requirement*

We can distinguish two types of information security:
(1) Personal identifiable information security, and
(2) Non-identifiable information security.
While the security requirements of NII are concerned with the traditional system characteristics of confidentiality, integrity, and availability, PII security lends itself to unique techniques pertaining only to PII.

The process of protecting PII involves (1) protection of the identities of the proprietor and (2) protection of the non-identity portion of the PII.

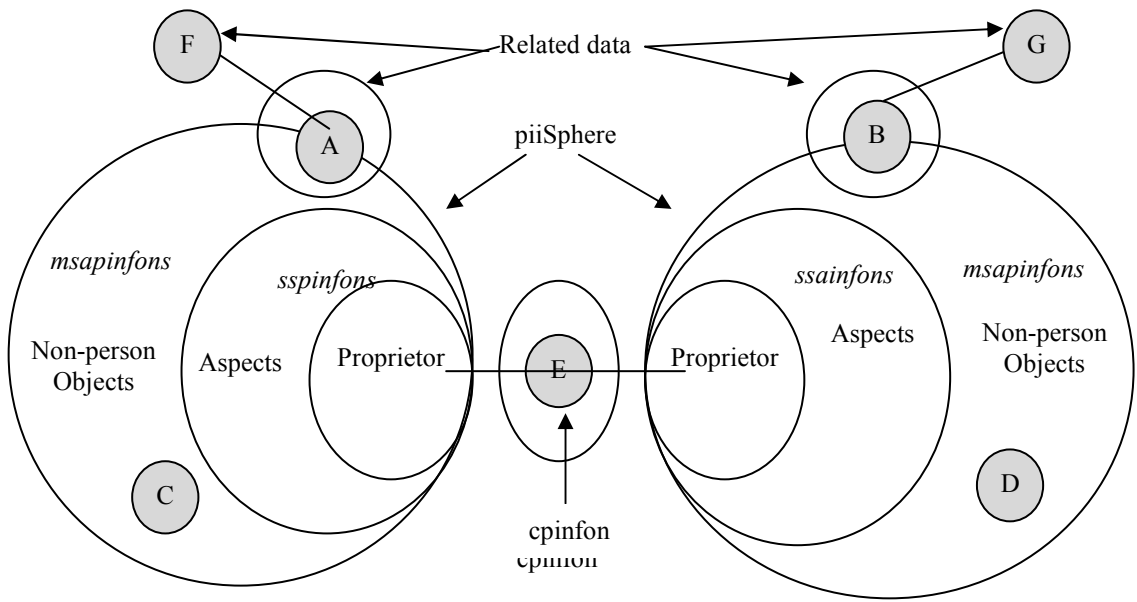

Figure 2. Conceptualization of piiSpheres of two proprietors.

Of course, all information security tools such as encryption can be applied in this context, yet other methods (e.g., anonymization) utilizing the unique structure of PII as a combination of identities and other information can also be used. Data-mining attacks on PII aim to determine the identity of the proprietor(s) from non-identifiable information; for example, determining the identity of a patient from anonymized information that gives age, sex, and zip code in health records (k-anonymization). Thus, PII lends itself to unique techniques that can be applied in protection of this information

Another important issue that motivates organizing PII separately is that any intrusion on PII involves information in addition to the owner's information (e.g., a company, proprietors, and other third parties, e.g., privacy commissioner). For example, a PII security system may require immediately alerting the proprietor that intrusion on his/her PII has occurred.

An additional point is that the sensitivity of PII is in general valued more highly than the sensitivity of other types of information. PII is more "valuable" than non-PII because of its privacy aspect, as discussed previously. Such considerations imply a special security status for PII. The source of this volubility is instigated by moral considerations [7].

*B. Policy requirement*

Some policies applied to PII are not applicable to NII (e.g., consent, opt-in/out, proprietor's identity management, trust, privacy mining). While NII security requirements are concerned with the traditional system characteristics of confidentiality, integrity, and availability, PII privacy requirements are also concerned with such issues as purpose, privacy compliance, transborder flow of data, third party disclosure, etc. Separating PII from NII can reduce the complex policies required to safeguard sensitive information where multiple rules are applied, depending on who is accessing the data and what the function is.

In general, PIIDB goes beyond mere protection of data:
1. PIIDB identifies proprietor's piiSphere and provides security, policy, and tools *to the piiSphere*.
2. PIIDB provides security, policy, and tools <u>only</u> *to proprietor's piiSphere*, thus conserving privacy efforts.
2. PIIDB identifies inter-piiSphere relationships (proprietors' relationships with each other) and provides security, policy, and tools to protect the privacy of these relationships.

## VII. PERSONAL IDENTIFIABLE INFORMATION DATABASE (PIIDB)

The central mechanism in PIIDB is an explicit declaration of proprietors in a table called PROPRIETORS that includes unique identifiers of all proprietors in the PIIDB. PROPRIETOR_TABLE contains a unique entry with an internal key (#proprietor) for each proprietor in addition to other information such as pointer(s) to his/her piiSphere.

The principle of uniqueness of proprietor's identifiers requires that the internal key (#proprietor) is mapped one-to-one to the individual's legal identity or physical location. This is an important feature in PIIDB to guarantee consistency of information about persons. This "identity" uniquely identifies the piiSphere and distinguishes one piiSphere from another. Thus, if we have PIIDB of three individuals, then we have three entries such that each leads (denoted as $\Rightarrow$) to three piiSpheres:

PROPRIETOR_TABLE:
$\{(\#proprietor1, \ldots) \Rightarrow$ piiSphere of proprietor 1, $(\#proprietor2, \ldots) \Rightarrow$ piiSphere of proprietor 2, $(\#proprietor3, \ldots) \Rightarrow$ piiSphere of proprietor 3$\}$.

The "..." denotes the possibility of other information in the table. What is the content of each piiSphere? The answer is set(s) of atomic PIIs and related information.

Usually, database design begins by identifying data items, including objects and attributes (Employee No., Name, Salary, Birth, Date of Employment, etc.). Relationships among data items are then specified (e.g., data dependencies). Semantically oriented graphs (e.g., ER graphs [13]) are sometimes used at this level. Finally, a set of tables is declared, such as the following:

T1 = Father (ID, Name, Details),
T2 = Mother (ID, Name Details),
T3 = Child (ID, Name, Details),
T4 = Case (No., Father ID, Mother ID, Child ID).

T1, T2, and T3 represent atomic PIIs of fathers, mothers, and children, respectively. T4 embeds compound PIIs. In PIIDB, if R is a compound PII, then it is represented by the set of atomic PIIs:

$\{R' =$ Case (No., Father ID),

$R'' =$ Case (No., Mother ID),

$R'' =$ Case (No., Child ID)$\}$

Where R' is in the piiSphere of father, R'' is in the piiSphere of mother, and R'' is in the piiSphere of child. Such a schema permits complete isolation of atomic PIIs from each other. This privacy requirement is essential in many personal identifiable databases. For example, in orphanages it is possible not to allow access to the information that a record exists in the database for a mother. In the example above, access policy for the three piiSpheres is independent from each other. At the conceptual level, reconstructing the relations among proprietors (cpinfons) is a database design problem (e.g., internal pointers among tables across piiSpheres).

PIIDB obeys all propositions defined previously. Some of these propositions can be utilized as privacy rules. As an illustration of the applications of these propositions, consider the case of privacy constraint that prohibits disclosing $\sigma \in$ PII. By proposition (9) above, mixing (e.g., amending, inserting, etc.) $\sigma$ with any other piece of information makes the disclosure constraint apply to the combined piece of information. In this case a general policy is: Applying a protection rule to σ1 ∈ PII implies applying the same protection to (σ₁⊠σ₂) where σ₂ ∉ PII.

VIII. CONCLUSION

The theory of PII infons can provide a theoretical foundation for technical solutions to problems of protection of personal identifiable information. In such an approach, privacy rules form an integral part of the design of the system. PII can be identified (hence becomes an object of privacy rules) during processing of information that may mix it with other types of information. Different types of basic PII infons provide an opportunity for tuning the design of an information system. We propose analyzing and processing PII as a database with clear boundary lines separate from non-identifiable information, which facilitates meeting the unique requirements of PII. A great deal of work is needed at the theoretical and design levels. An expanded version of this paper includes complete formalization of the theory. Additionally, we are currently applying the approach to analysis of an actual database of a government agency that handles social problems where a great deal of PII is collected.

AUTHOR PROFILES

**Sabah Al-Fedaghi** holds an MS and a PhD in computer science from Northwestern University, Evanston, Illinois, and a BS in computer science from Arizona State University, Tempe. He has published papers in journals and contributed to conferences on topics in database systems, natural language processing, information systems, information privacy, information security, and information ethics. He is an associate professor in the Computer Engineering Department, Kuwait University. He previously worked as a programmer at the Kuwait Oil Company and headed the Electrical and Computer Engineering Department (1991–1994) and the Computer Engineering Department (2000–2007).

**Bernhard Thalheim** holds an MSc in mathematics from Dresden University of Technology, a PhD in Mathematics from Lomonosov University Moscow, and a DSc in Computer Science from Dresden University of Technology. His major research interests are database theory, logic in databases, discrete mathematics, knowledge systems, and systems development methodologies, in particular for Web information systems. He has been program committee chair and general chair for several international events, including ADBIS, ASM, EJC, ER, FoIKS, MFDBS, NLDB, and WISE. He is currently full professor at Christian-Albrechts-University at Kiel in Germany and was previously with Dresden University of Technology (1979–1988) (associate professor beginning in 1986), Kuwait University (1988–1990) (visiting professor), Rostock University (1990–1993) (full professor), and Cottbus University of Technology (1993–2003) (full professor).